# Quantum Whistling in Superfluid $^4$He

E. Hoskinson and R. E. Packard

Physics Department, University of California, Berkeley, California 94720, USA

**Fundamental considerations predict that macroscopic quantum systems such as superfluids and the electrons in superconductors will exhibit oscillatory motion when pushed through a small constriction. Here we report the observation of these oscillations between two reservoirs of superfluid $^4$He partitioned by an array of nanometer-sized apertures. They obey the Josephson frequency equation and are coherent amongst all the apertures. This discovery at the relatively high temperature of 2K (2000 times higher than related phenomena in $^3$He) may pave the way for a new class of practical rotation sensors of unprecedented precision.**

The Josephson effects in superconductors have received enormous attention both for scientific understanding and for technological importance[1]. Many analogous effects, including Josephson oscillations, have been observed[2,3] in superfluid $^3$He below $10^{-3}$K. However, detection of oscillations at the Josephson frequency in superfluid $^4$He has remained elusive until now despite almost four decades of attempts[4].

Superconductors and superfluids are both described by a macroscopic wave function which includes both amplitude and phase, $\phi$. A chemical potential difference $\Delta\mu = \mu_2 - \mu_1$ between two baths of superfluid separated by an aperture causes the phase difference $\Delta\phi = \phi_2 - \phi_1$ to change in accordance with the Josephson–Anderson phase evolution equation,

$$\frac{d\Delta\phi}{dt} = \frac{-\Delta\mu}{\hbar} \qquad \mathbf{1.}$$

Here $\hbar$ is Planck's constant $h$ divided by $2\pi$ and $\Delta\mu/m_4 = \Delta P/\rho - s\Delta T$ where $m_4$ is the mass of a $^4$He atom, $\Delta P$ is the pressure difference, $\rho$ is the mass density, $s$ is the entropy per unit mass, and $\Delta T$ is the temperature difference. A non-zero $\Delta\phi$ results in a superfluid current $I(\Delta\phi)$ through the aperture. If $I(\Delta\phi)$ is $2\pi$ periodic, a constant $\Delta\mu$ causes current to oscillate through the aperture at the Josephson frequency $f_j = \Delta\mu/h$. The periodicity in $I(\Delta\phi)$ can occur if the aperture acts like an ideal weak link[3,5], in which case $I(\Delta\phi) \propto \sin(\Delta\phi)$, or by the generation of $2\pi$ phase slips[6], in which case $I(\Delta\phi)$ is expected to follow a sawtooth waveform.

A schematic of the experiment is shown in figure 1a. We use an electrostatically driven diaphragm[2] to apply an initial pressure step between two baths of superfluid separated by an aperture array. The array is 65x65 nominally 70nm apertures spaced on a 3μm square lattice in a 50nm thick silicon-nitride membrane. Following the pressure step, fluid flows through the array and the chemical potential difference relaxes to zero. When the output of a diaphragm position sensor monitoring fluid flow is connected to a set of headphones, we hear a clear whistling sound passing from high to low frequency.

By Fourier transform methods we extract the frequency and amplitude of this whistle as a function of time throughout the transient. Immediately after the pressure step is applied, the temperatures on either side of the aperture array are equal and the entire $\Delta\mu$ is determined by the initial pressure head, $\Delta P_o$. Figure 1b shows that the initial frequency is accurately proportional to the initial chemical potential difference and the slope of the line agrees, within the systematic error of our pressure calibration, with the Josephson frequency formula, $f_j = m_4 \Delta P_o / \rho h$.

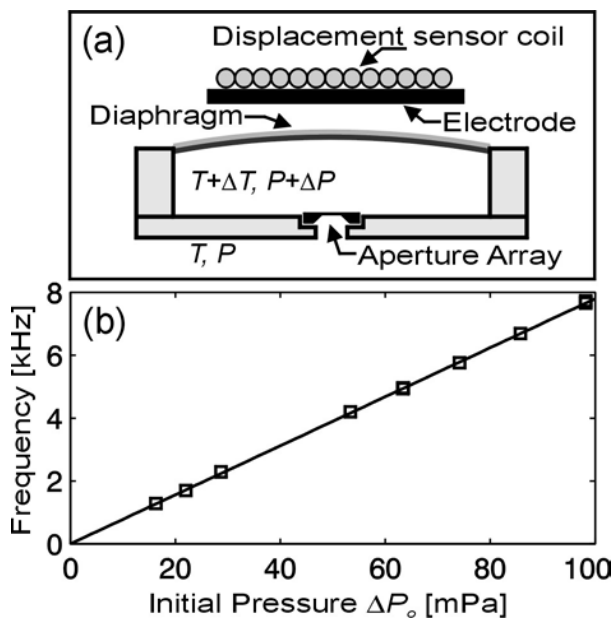

**Figure 1.** *Quantum oscillations in $^4$He. (a) Schematic of the experimental cell, described in more detail in the supplemental material. T. Haard made important contributions to the cell design and construction. (b) Plot of the whistle frequency vs. the initial pressure, $\Delta P_o = \rho \Delta \mu_o / m_4$. Temperature is in the range $T_\lambda - T = 1.7mK$ to $2.9mK$, where $T_\lambda$ is the superfluid transition temperature. A fit (solid line) to the data (squares) gives a slope of 78 Hz/mPa, with a systematic uncertainty of 20% arising from our pressure calibration. This is in agreement with the Josephson frequency relation $f_j = \Delta\mu/h$ value of 68.7 Hz/mPa. The oscillation is still present down to at least 150mK below $T_\lambda$. In this regime the healing length is much smaller than the aperture diameter and $I(\Delta\phi)$ is linear. Presumably the oscillation is due to periodic $2\pi$ phase slips.*

Oscillations resulting from $2\pi$ phase slips are expected to have a velocity amplitude $\kappa/2\ell$ where $\kappa = h/m_4$ is the circulation quantum and $\ell$ is an effective length for one aperture[7]. If, in addition, the oscillation in each of the $N$ apertures occurs coherently, the amplitude of the diaphragm displacement Fourier component at $f_j$ is

$$X_o = \alpha \frac{\rho_s N \kappa a}{4\pi f_j \rho A \ell}, \quad 2.$$

where $A$ is the area of the diaphragm, $a$ is the area of a single aperture, and $\rho_s$ is the superfluid density. The factor $\alpha$ would be $2/\pi$ for a sawtooth waveform, or unity for a sinusoid of the same peak amplitude. We find $\alpha \sim 0.6$, independent of temperature in the range $T_\lambda$-$T$ = 1.7mK to 2.9mK. Thus we conclude that the oscillation is a coherent phenomenon involving all the apertures in the array, possibly sawtooth in waveform. This coherence is remarkable, since earlier work using a single aperture showed that thermal fluctuations in the phase slip nucleation process destroy time coherence in the rate of phase slippage so that no Josephson oscillation exists[8]. It seems thermal fluctuations are suppressed for an array[9]. This is an intriguing observation deserving further study.

We have found that superfluid $^4$He in an array of small apertures behaves quantum coherently, oscillating at the Josephson frequency. Since these oscillations appear in $^4$He at 2000 times higher temperature than in superfluid $^3$He, the possibility is raised that sensitive rotation sensors can be built using much simpler technology than heretofore thought possible[10,11,12,13]. Such sensors will find utility in rotational seismology, geodesy and possibly tests of general relativity.

This work was supported in part by NSF grant #DMR-0244882 and NASA.

**Supplementary Information**

We use a single diaphragm cell shown in Figure 1. Our aperture array is a 65x65 square lattice with 3μm spacing, of nominally 70nm diameter holes formed in a 50nm thick silicon nitride membrane[1]. By applying an electrostatic voltage between the metalized diaphragm and an adjacent electrode we create an initial chemical potential difference $\Delta\mu$ across the aperture array. The diaphragm's motion as the system relaxes is detected with a sensitive displacement transducer[2].

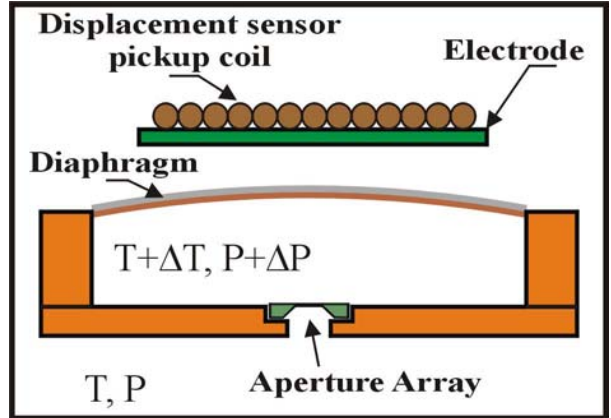

*Figure 1. Schematic of the experimental cell. A cylindrical washer-shaped spacer of height 0.6mm and inner diameter 8mm is bounded on the top by a flexible Kapton diaphragm and on the bottom by a rigid plate containing the aperture array. The array is produced using e-beam lithography in a 50nm thick 200μm x 200μm silicon nitride membrane supported by a silicon frame which is glued into the lower plate. The top surface of the Kapton is coated with lead. Its position, determined by an adjacent superconducting displacement sensor[2], is proportional to $\Delta P$ and acts as a pressure gauge. We initially calibrate this gauge using a fountain pressure technique to be described elsewhere. The cell sits inside a metal can which is immersed in a pumped bath dewar of liquid helium. The can and cell are filled with $^4$He through a cryogenic valve. The temperature of the bath and the $^4$He inside the can is controlled by a standard feedback loop.*

Figure 2 shows a typical relaxation transient. It consists of an initial highly dissipative flow followed by a characteristic lightly damped Helmholtz oscillation. It is only during the pre-Helmholtz oscillation part of the transient that the oscillations occur. In this region $\Delta\mu$ is non-zero throughout, but continuously drops as the system relaxes. If the output of the transducer is connected to a set of headphones while $\Delta\mu$ relaxes, we hear a clear whistling sound passing from high to low frequency.

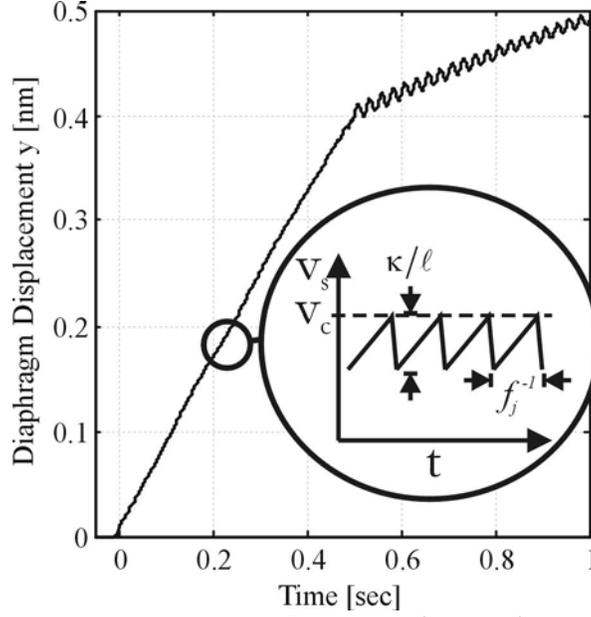

*Figure 2. A typical transient. At time t=0, we apply a voltage between the metalized diaphragm and a nearby electrode (Figure 1), pulling on the diaphragm and establishing a pressure difference $\Delta P_0$. Fluid flows through the aperture array and the diaphragm position, y, relaxes toward a new equilibrium position, $y_{eq}$. The pressure difference is proportional to the displacement from this new equilibrium: $\Delta P \propto (y-y_{eq})$. We are interested in the flow that precedes the large amplitude Helmholtz oscillations which start at t ~ 0.5sec. In the inset, we show what a noise-free trace of the superfluid velocity $v_s$ might look like if we were to zoom in on a short section of this flow. Note that $v_s$ is proportional to $\partial y/\partial t$. This sawtooth pattern with amplitude $\kappa/2\ell$ would be expected for a linear current phase relation and $2\pi$ phase slips. The frequency of the waveform is equal to the Josephson frequency $f_j = \Delta\mu/h$. It reaches zero where the Helmholtz mode begins. During the Helmholtz mode $\Delta\mu$ oscillates about zero, and $\Delta P$, which has not yet reached zero, continues to relax. This transient was taken at $T_\lambda - T = 2.9mK$.*

In an "ideal" Josephson junction, or weak link, the current-phase relation is sinusoidal: $I(\Delta\phi) \propto \sin(\Delta\phi)$. One requirement for such a weak link is that its size must be comparable to or smaller than the coherence length (for a superconductor), or healing length (for a superfluid). If the healing length is instead much smaller that the aperture, then in $^4$He the current phase relation is expected to be linear and phase slips are possible. The healing length for $^4$He is a strong function of the temperature near the superfluid transition temperature $T_\lambda$ and is expected to follow,

$$\xi_4 = \frac{\xi_o}{(1-T/T_\lambda)^\nu} \qquad 1,$$

where $\xi_o \approx 0.3nm$, $T_\lambda = 2.176\,K$ and $\nu = 0.67$, although estimates of $\xi_o$ vary by as much as a factor of 3. Using the above formula, at $T_\lambda - T = 1mK$, $\xi_o \approx 60nm$, and at $T_\lambda - T = 100mK$, $\xi_o \approx 3nm$. Therefore one would expect that for temperatures closer than 1mK to $T_\lambda$ our apertures might act like ideal weak links and exhibit a sinusoidal current-phase relation, but that for temperatures farther away from $T_\lambda$, the current-phase relation would be linear. Flow features in a hydrodynamic resonator at $T_\lambda - T \leq 60\mu K$ have been found to be consistent with a sine-like current phase relation[3] in $^4$He.

In this work we were mainly limited to the temperature regime a few mK below the transition temperature $T_\lambda$ for the following reasons. Measurements closer to $T_\lambda$ require better temperature stability than we have at present. Since the critical velocity, $v_c$, increases as temperature drops[4], at lower temperatures $f_j$ sweeps from a maximum to zero so quickly that we cannot discern discrete frequencies in the relaxation transients. Using a dc-drive technique not described here we have observed the coherent oscillations as low as 150mK below $T_\lambda$, where $I(\Delta\phi)$ is expected to be purely linear. The measured oscillation amplitude also supports the idea that the current-phase relation is not purely sinusoidal.

The expression for the Fourier component of the diaphragm displacement at the Josephson frequency $f_j$ is

$$X_o = \alpha \frac{\rho_s N \kappa a}{4\pi f_j \rho A \ell}.$$

This formula follows from the expected velocity amplitude $\kappa/2\ell$, which, with $N$ apertures each of area $a$, gives rise to a total superfluid mass current amplitude $\rho_s N \kappa a / 2\ell$ oscillating through the aperture array. This oscillating current causes the diaphragm to oscillate with velocity amplitude $\rho_s N \kappa a / \rho A 2\ell$, where $A$ is the area of the diaphragm. If the oscillation has a sawtooth waveform, the magnitude of its first Fourier harmonic is $2/\pi$ times the amplitude of the sawtooth, whereas if the oscillation is a sinusoid, this factor is unity. Finally, conversion of the diaphragm velocity amplitude to the stated displacement amplitude is achieved by dividing the velocity amplitude by $2\pi f_j$.

We have shown that an oscillation obeying the Josephson frequency relation in $^4$He exists and have provided strong evidence that it occurs in a quantum coherent fashion amongst all the apertures in the array. The exact form of the current-phase relation and how it varies with temperature, including the expected crossover from linear to sinusoidal as $T$ approaches $T_\lambda$, remains an intriguing problem.